# Quantum Phase Space Approach to the Ideal Fermi and Bose Gases

**Rivo Herivola Manjakamanana Ravelonjato[1], Ravo Tokiniaina Ranaivoson[1], Raoelina Andriambololona[1], Roland Raboanary[2], Naivo Rabesiranana[1, 2], Solofoarisina Wilfrid Chrysante[1].**

[1] Institut National des Sciences et Techniques Nucléaires (INSTN- Madagascar)
*BP 3907 Antananarivo 101, Madagascar, instn@moov.mg*

[2] Department of Physics, Faculty of Sciences – University of Antananarivo
*BP 566 Antananarivo 101, Madagascar*

email: *manjakamanana@yahoo.fr[1]*

**Abstract**:

In this work, improvements are introduced to the current models of the ideal Fermi gas and the ideal Bose gas by incorporating the quantum nature of phase space, which is directly linked to the uncertainty principle. These improved models build upon the recently developed concepts of quantum phase space (QPS) and the QPS representation of quantum mechanics. The Hamiltonian operator for a gas particle and its eigenstates are first determined, and quantum statistical mechanics is used to derive the thermodynamic properties of the ideal gas.

Analytic expressions for thermodynamic quantities—including the grand canonical potential, particle number, internal energy, von Neumann entropy, and pressure—are derived, along with the corresponding thermodynamic equations of state for both bosons and fermions. These corrections are particularly significant at low temperatures and in confined volumes, where quantum effects related to system geometry, such as shape and size, become significant. The results also establish a direct link between thermodynamic functions and the quantum statistical variances of momentum.

Importantly, the improved models recover well-known classical relations of the ideal gas in the high-temperature and large-volume limits, ensuring consistency with classical physics. By addressing quantum corrections and their thermodynamic implications, this work provides a foundation for further applications in nanoscale systems, quantum gases, low-temperature physics, ultracold physics, and astrophysics.

Keywords : Quantum phase space, Quantum statistical mechanics, Anisotropic pressure, Momentum variance matrix, Confinement effects.

# 1. Introduction

Since the advent of modern physics and technological advances, the validity of classical thermodynamics has been questioned. To address this, various methodologies have been explored to incorporate quantum and relativistic corrections into statistical mechanics and thermodynamics, as discussed in [1-14]. One essential approach to linking statistical mechanics and thermodynamics is through the concept of phase space. In classical physics, phase space is intuitively defined as the set of all possible simultaneous values of canonical coordinates and momenta. However, extending this definition within the framework of quantum physics is challenging due to the uncertainty principle [15–19].

Various efforts have been made to address the challenges associated with the concept of phase space in quantum physics. One prominent approach is the phase space formulation of quantum mechanics, which employs Wigner quasi-probability distributions [1, 20-24]. This framework has provided significant insights into quantum systems but also yields counterintuitive results, such as negative values in the Wigner function. These negative values, which stem from the use of classical phase space concepts to represent quantum phenomena, limit its interpretation as a genuine probability density function. To overcome these challenges, alternative methods have been proposed to refine the quantum phase space (QPS) formulation [25-27]. Additionally, it has been shown in [28], that the quantum partition function of an ideal gas can be computed without relying on the phase space framework. Instead, this is achieved by summing Boltzmann factors over quantum states of a particle in a box.

Among the various approaches to QPS, the one introduced and developed in refs. [16–19] stands out as fundamentally different from the earlier approaches, offering a unique perspective. Building on this framework, the same group of authors demonstrated in [29] that the concept of QPS enables the derivation of quantum and relativistic corrections to the Maxwell-Boltzmann ideal gas model. These corrections are fundamentally associated with the quantum nature of phase space and are directly related to the uncertainty principle. Motivated by the relevance of these findings, this study extends the approach to ideal Fermi and Bose gases. Our study aims to improve these models by determining the quantum corrections introduced through the use of a QPS.

The statistical description of an ideal quantum gas, consistent with the concept of QPS, is presented in Section 2. Then the grand canonical potentials corresponding to bosons and fermions are determined in Section 3, along with the relation between the momenta variances and the thermodynamic variables. In Section 4, the explicit expressions for thermodynamic quantities such as particle number, internal energy, pressure, and equations of state are deduced from the grand canonical potentials. Section 5 corresponds to the discussion and conclusion. BBoldfaced letters are used for quantum operators, while regular letters denote eigenvalues.

# 2. Statistical description of an ideal gas using the concept of quantum phase space

## 2.1 Hamiltonian operators, number operator and their eigenstates

The concept of QPS was introduced, developed, and applied to ideal gas models in the references [16-19, 29]. Some results from these references are used in the present work.

Let us consider a particle $A$ in an ideal gas. As shown in [18, 29], within the phase space representation of quantum mechanics, the Hamiltonian operator for this particle $A$ in its mean rest frame is

$$\boldsymbol{h_A} = \sum_{l=1}^{3} (2\boldsymbol{z}_{Al}^{\dagger}\boldsymbol{z}_{Al} + 1)\frac{\mathcal{B}_{ll}}{m} \tag{1}$$

where $\boldsymbol{z}_{Al}$ and $\boldsymbol{z}_{Al}^{\dagger}$ are ladder operators related to the coordinates and momenta operators of the particle**,** $l$ is the index related to space direction ($l = 1,2,3$), $m$ is the mass of the particle $A$ and $\mathcal{B}_{ll}$ is the ground momentum variance in the $l$-$th$ direction of the three dimensional space.
The eigenvalue equations of the hamiltonian operator $\boldsymbol{h}_A$ is

$$\boldsymbol{h_A}|\alpha_A\rangle = \varepsilon_A|\alpha_A\rangle \tag{2}$$

$|\alpha_A\rangle$ is the eigenstate and $\varepsilon_A$ is the corresponding eigenvalue. The explicit expression of the eigenvalue $\varepsilon_A$ is [18, 29]:

$$\varepsilon_A = \sum_{l=1}^{3} (2n_{Al} + 1)\frac{\mathcal{B}_{ll}}{m} = (2n_{A1} + 1)\frac{\mathcal{B}_{11}}{m} + (2n_{A2} + 1)\frac{\mathcal{B}_{22}}{m} + (2n_{A3} + 1)\frac{\mathcal{B}_{33}}{m} \tag{3}$$

in which $n_{A1}, n_{A2}, n_{A3}$ are positive integers. Taking account of the relation (3), we can express the eigenstate as $|\alpha_A\rangle$

$$|\alpha_A\rangle = |n_{A1}, n_{A2}, n_{A3}\rangle \tag{4}$$

The Hamiltonian operator $\boldsymbol{H}_G$ of an ideal gas composed of $N_G$ particles can be deduced using (1) and by summing over all particles.

$$\boldsymbol{H}_G = \sum_{A=1}^{N_G} \boldsymbol{h_A} = \sum_{A=1}^{N_G}\sum_{l=1}^{3} \left(2\boldsymbol{z}_{Al}^{\dagger}\boldsymbol{z}_{Al} + 1\right)\frac{\mathcal{B}_{ll}}{m} \tag{5}$$

For bosonic and fermionic particles, there is no need to assign the "$A$" index to a particular particle since the particles are all indistinguishable from each other. Therefore, we can simply write for a particle energy eigenstate (instead of (4))

$$|\alpha\rangle = |n_1, n_2, n_3\rangle \tag{6}$$

An eigenstate $|G\rangle$ of the Hamiltonian operator $\boldsymbol{H}_G$ can be deduced from the one particle energy eigenstates $|\alpha\rangle$ as symmetrized Fock state (for bosons) or antisymmetrized Fock state (for fermions). It can be written in the form

$$|G\rangle = |N(0,0,0); N(1,0,0); N(0,1,0), N(0,0,1); \dots; N(n_1, n_2, n_3); \dots\rangle \tag{7}$$

in which $N_\alpha = N(n_1, n_2, n_3)$ represents the number of particles in a particular state $|\alpha\rangle = |n_1, n_2, n_3\rangle$. We have the eigenvalue equation for the Hamiltonian operator $\boldsymbol{H}_G$ of the gas:

$$\boldsymbol{H}_G|G\rangle = E_G|G\rangle \tag{8}$$

with the eigenvalue

$$E_G = \sum_{\alpha} N_\alpha \varepsilon_\alpha = \sum_{n_1} \sum_{n_2} \sum_{n_3} N(n_1, n_2, n_3)\varepsilon(n_1, n_2, n_3) \tag{9}$$

in which $\varepsilon_\alpha = \varepsilon(n_1, n_2, n_3)$ is the energy value corresponding to a state $|\alpha\rangle = |n_1, n_2, n_3\rangle$

$$\varepsilon_\alpha = \varepsilon(n_1, n_2, n_3) = \sum_{l=1}^{3} (2n_l + 1)\frac{\mathcal{B}_{ll}}{m} = (2n_1 + 1)\frac{\mathcal{B}_{11}}{m} + (2n_2 + 1)\frac{\mathcal{B}_{22}}{m} + (2n_3 + 1)\frac{\mathcal{B}_{33}}{m} \tag{10}$$

We may consider the operator $\boldsymbol{N}_G$ associated to the total number of particles $N_G$ in the gas. $\boldsymbol{H_G}$ and $\boldsymbol{N_G}$ have the same eigenstate $|G\rangle$. Eigenvalues equation of $\boldsymbol{N}_G$ is

$$\boldsymbol{N}_G|G\rangle = N_G|G\rangle \tag{11}$$

We introduce the operator $\boldsymbol{N_\alpha}$ associated with the particles number in a particular state $|\alpha\rangle = |n_1, n_2, n_3\rangle$ . The corresponding eigenvalue equation is

$$\boldsymbol{N_\alpha}|G\rangle = N_\alpha|G\rangle = N(n_1, n_2, n_3)|N(0,0,0); \dots; N(n_1, n_2, n_3); \dots\rangle \tag{12}$$

with $N_\alpha = N(n_1, n_2, n_3)$ the eigenvalue of the operator $\boldsymbol{N_\alpha}$ **:** it is the total number of particles in the state $|\alpha\rangle = |n_1, n_2, n_3\rangle$.

We have the following relations between the operators $\boldsymbol{H}_G, \boldsymbol{N}_G$ and $\boldsymbol{N_\alpha}$:

$$\begin{cases} \boldsymbol{H}_G = \sum_{\alpha} \boldsymbol{N}_\alpha \varepsilon_\alpha \\ \boldsymbol{N}_G = \sum_{\alpha} \boldsymbol{N_\alpha} \end{cases} \tag{13}$$

We obtain, between the eigenvalues, the following relations

$$\begin{cases} E_G = \sum_{\alpha} N_\alpha \varepsilon_\alpha \\ N_G = \sum_{\alpha} N_{\boldsymbol{\alpha}} \end{cases} \tag{14}$$

## 2.2 Quantum gas in thermal equilibrium with an external heat bath

At thermodynamic equilibrium, the interaction between a quantum gas and an external heat bath can be described using the grand canonical ensemble:

- A possible microstate is the quantum state $|G\rangle$ given explicitly by the relation (9). A state $|G\rangle$ is defined by the exact values of the energy $E_G$ and the particles number $N_G$ of the gas. The expressions of $E_G$ and $N_G$ are given by the relation (14)

- A macrostate is defined by its thermodynamic internal energy $U$ and particles number**.** With these main quantities are associated the thermodynamics variables like the temperature $T$, the pressure $P$ and the volume $V$. In the framework of statistical quantum mechanics, this macrostate is described by the density operator $\boldsymbol{\rho}$ :

$$\boldsymbol{\rho} = \frac{1}{\Xi} e^{-\beta(\boldsymbol{H}_G - \mu \boldsymbol{N}_G)} = \frac{1}{\Xi} e^{-\frac{1}{k_B T}(\boldsymbol{H}_G - \mu \boldsymbol{N}_G)} \tag{15}$$

in which $\Xi$ is the grand canonical partition function, $\boldsymbol{H}_G$ and $\boldsymbol{N}_G$ are respectively the Hamiltonian and the total number of particles operators which are given in the relation (13). When the gas is at thermodynamic equilibrium with a heat bath, its macrostate remains the same while its microstates may change.

The probability $P_G$ that the microstate of the gas takes a particular value $|G\rangle$ is equal to the corresponding eigenvalue of the density operator $\boldsymbol{\rho}$ given in (15)

$$\begin{cases} P_G = \dfrac{1}{\Xi} e^{-\beta(E_G - \mu N_G)} \\ \boldsymbol{\rho}|G\rangle = P_G|G\rangle \end{cases} \tag{16}$$

$E_G$ and $N_G$ are the eigenvalues of operators $\boldsymbol{H}_G$ and $\boldsymbol{N}_G$ corresponding to relations (14). The normalization of the probabilities $P_G$, which is equivalent to the relation $Tr(\boldsymbol{\rho}) = 1$, leads to the expression of the grand canonical partition function $\Xi$.

$$\sum_{\{|G\rangle\}} P_G = 1 \Leftrightarrow \Xi = Tr\left[e^{-\beta(\boldsymbol{H}_G - \mu \boldsymbol{N}_G)}\right] = \sum_{\{|G\rangle\}} e^{-\beta(E_G - \mu N_G)} \tag{17}$$

The summation (17) is performed on the set of microstates $\{|G\rangle\}$ which corresponds to the same thermodynamic macrostate of the gas at equilibrium. This macrostate is defined by the internal energy $U = E = \langle E_G \rangle$ and by the thermodynamic particles number $N = \langle N_G \rangle$ which are given by the following relations

$$\begin{cases} U = \langle E_G \rangle = Tr(\boldsymbol{\rho}\boldsymbol{H}_G) = \displaystyle\sum_{\{|G\rangle\}} P_G E_G = \frac{1}{\Xi} \sum_{\{|G\rangle\}} E_G e^{-\beta(E_G - \mu N_G)} \\ N = \langle N_G \rangle = Tr(\boldsymbol{\rho}\boldsymbol{N}_G) = \displaystyle\sum_{\{|G\rangle\}} P_G N_G = \frac{1}{\Xi} \sum_{\{|G\rangle\}} N_G e^{-\beta(E_G - \mu N_G)} \end{cases} \tag{18}$$

All thermodynamic variables of a gas can be derived from the partition function $\Xi$ or from the grand canonical potential $\Phi_G$. $\Phi_G$ is related to the grand canonical partition function $\Xi$ by the following relation :

$$\Phi_G = -k_B T ln(\Xi) = -\frac{1}{\beta} ln(\Xi) \tag{19}$$

### 2.3 Partition functions of states and associated particle mean numbers

The expression of the partition function $\Xi$ of the quantum gas can be deduced from the relations (14) and (17):

- For the case of bosons, $N_\alpha$ takes any integer values so we obtain

$$\Xi = \sum_{\{|G\rangle\}} e^{-\beta(E_G-\mu N_G)} = \prod_{\{|\alpha\rangle\}} \sum_{N_\alpha=0}^{\infty} e^{-\beta N_\alpha(\varepsilon_\alpha-\mu)} \tag{20}$$

- For the case of fermions, $N_\alpha$ takes only the value 0 or 1 because of the Pauli exclusion principle, so we obtain

$$\Xi = \sum_{\{|G\rangle\}} e^{-\beta(E_G-\mu N_G)} = \prod_{\{|\alpha\rangle\}} \sum_{N_\alpha=0}^{1} e^{-\beta N_\alpha(\varepsilon_\alpha-\mu)} \tag{21}$$

For every case, $\Xi$ can be put in the form

$$\Xi = \prod_{\{|\alpha\rangle\}} \Xi_\alpha \tag{22}$$

with $\Xi_\alpha$ the grand canonical partition function associated to the state $|\alpha\rangle = |n_1, n_2, n_3\rangle$

$$\begin{cases} \Xi_\alpha = \sum_{N_\alpha=0}^{\infty} e^{-\beta(\varepsilon_\alpha-\mu)N_\alpha} = \dfrac{1}{1-e^{-\beta(\varepsilon_\alpha-\mu)}} & \text{for bosons} \\ \Xi_\alpha = \sum_{N_\alpha=0}^{1} e^{-\beta(\varepsilon_\alpha-\mu)N_\alpha} = 1+e^{-\beta(\varepsilon_\alpha-\mu)} & \text{for fermions} \end{cases} \tag{23}$$

$\varepsilon_\alpha$ is the energy level corresponding to the state $|\alpha\rangle = |n_1, n_2, n_3\rangle$ and $N_\alpha = N(n_1, n_2, n_3)$ corresponds to the particles number corresponding to the state $|\alpha\rangle$.

The probability $P_\alpha$ to have $N_\alpha$ particles in the state $|\alpha\rangle$ is

$$\begin{cases} P_\alpha = \dfrac{e^{-\beta(\varepsilon_\alpha-\mu)N_\alpha}}{\Xi_\alpha} = e^{-\beta(\varepsilon_\alpha-\mu)N_\alpha}\left(1-e^{-\beta(\varepsilon_\alpha-\mu)}\right) & \text{for bosons} \\ P_\alpha = \dfrac{e^{-\beta(\varepsilon_\alpha-\mu)N_\alpha}}{\Xi_\alpha} = \dfrac{e^{-\beta(\varepsilon_\alpha-\mu)N_\alpha}}{1+e^{-\beta(\varepsilon_\alpha-\mu)}} = \dfrac{1}{e^{\beta(\varepsilon_\alpha-\mu)N_\alpha}+1} & \text{for fermions} \end{cases} \tag{24}$$

We may deduce from (24) the mean value $\langle N_\alpha\rangle$ of the particles number in the particular state $|\alpha\rangle = |n_1, n_2, n_3\rangle$

$$\begin{cases} \langle N_\alpha\rangle = \sum_{N_\alpha=0}^{+\infty} N_\alpha P_\alpha = \left(1-e^{-\beta(\varepsilon_\alpha-\mu)}\right)\sum_{N_\alpha=0}^{+\infty} N_\alpha e^{-\beta(\varepsilon_\alpha-\mu)N_\alpha} = \dfrac{1}{e^{\beta(\varepsilon_\alpha-\mu)}-1} & \text{for bosons} \\ \langle N_\alpha\rangle = \sum_{N_\alpha=0}^{1} N_\alpha P_\alpha = \sum_{N_\alpha=0}^{1} \dfrac{N_\alpha}{e^{\beta(\varepsilon_\alpha-\mu)N_\alpha}+1} = \dfrac{1}{e^{\beta(\varepsilon_\alpha-\mu)N_\alpha}+1} & \text{for fermions} \end{cases} \tag{25}$$

The relations in (25) correspond respectively to the well-known Bose-Einstein and Fermi-Dirac distributions. If we consider the spin degeneracy $g_S$, we should write instead of (25) the following relations :

$$\begin{cases} \langle N_\alpha \rangle = \dfrac{g_S}{e^{\beta(\varepsilon_\alpha-\mu)}-1} & \text{for bosons} \\ \langle N_\alpha \rangle = \dfrac{g_S}{e^{\beta(\varepsilon_\alpha-\mu)N_\alpha}+1} & \text{for fermions} \end{cases} \quad (26)$$

## 3. Grand canonical potential function and momenta variances

### 3.1 Determination of the grand canonical potential function

The grand potential is given by the relation (19). Using the relation (22), we obtain:

$$\Phi_G = -\frac{1}{\beta} ln\left(\prod_{\{|\alpha\rangle\}} \Xi_\alpha\right) = -\frac{1}{\beta}\sum_{\{|\alpha\rangle\}} ln(\Xi_\alpha) \quad (27)$$

More explicitly, we have

$$\begin{cases} \Phi_G = \dfrac{1}{\beta}\displaystyle\sum_{\{|\alpha\rangle\}} ln\left(1-e^{-\beta(\varepsilon_\alpha-\mu)}\right) = -\dfrac{1}{\beta}\displaystyle\sum_{\{|\alpha\rangle\}}\sum_{j=1}^{+\infty}\dfrac{(\xi)^j e^{-j\beta\varepsilon_\alpha}}{j} & \text{for bosons} \\ \Phi_G = -\dfrac{1}{\beta}\displaystyle\sum_{\{|\alpha\rangle\}} ln\left(1+e^{-\beta(\varepsilon_\alpha-\mu)}\right) = \dfrac{1}{\beta}\displaystyle\sum_{\{|\alpha\rangle\}}\sum_{j=1}^{+\infty}\dfrac{(-\xi)^j e^{-j\beta\varepsilon_\alpha}}{j} & \text{for fermions} \end{cases} \quad (28)$$

with $\xi = e^{\beta\mu}$.

The explicit expression of the energy $\varepsilon_\alpha = \varepsilon(n_1, n_2, n_3)$ corresponding to a state $|\alpha\rangle = |n_1, n_2, n_3\rangle$ as given in the relation (10) can be put in the following form

$$\varepsilon_\alpha = \varepsilon(n_1, n_2, n_3) = 2n_1\frac{\mathcal{B}_{11}}{m} + 2n_2\frac{\mathcal{B}_{22}}{m} + 2n_3\frac{\mathcal{B}_{33}}{m} + \frac{\mathcal{B}_{11}}{m} + \frac{\mathcal{B}_{22}}{m} + \frac{\mathcal{B}_{33}}{m} \quad (29)$$

Introducing the relation (29) in the relation (28), we can write the expression of $\Phi_G$ in the following form:

$$\begin{cases} \Phi_G = -\dfrac{g_s}{\beta}\displaystyle\sum_{j=1}^{+\infty}\dfrac{(\xi)^j}{j}\sum_{n_1=0}^{+\infty}\sum_{n_2=0}^{+\infty}\sum_{n_3=0}^{+\infty}\left[e^{-j\beta\left(2n_1\frac{\mathcal{B}_{11}}{m}+2n_2\frac{\mathcal{B}_{22}}{m}+2n_3\frac{\mathcal{B}_{33}}{m}\right)}e^{-j\beta\left(\frac{\mathcal{B}_{11}}{m}+\frac{\mathcal{B}_{22}}{m}+\frac{\mathcal{B}_{33}}{m}-\mu\right)}\right] & \text{for bosons} \\ \Phi_G = \dfrac{g_s}{\beta}\displaystyle\sum_{j=1}^{+\infty}\dfrac{(-\xi)^j}{j}\sum_{n_1=0}^{+\infty}\sum_{n_2=0}^{+\infty}\sum_{n_3=0}^{+\infty}\left[e^{-j\beta\left(2n_1\frac{\mathcal{B}_{11}}{m}+2n_2\frac{\mathcal{B}_{22}}{m}+2n_3\frac{\mathcal{B}_{33}}{m}\right)}e^{-j\beta\left(\frac{\mathcal{B}_{11}}{m}+\frac{\mathcal{B}_{22}}{m}+\frac{\mathcal{B}_{33}}{m}-\mu\right)}\right] & \text{for fermions} \end{cases} \quad (30)$$

with $g_S$ is the spin degeneracy. The summation over the set of the integers $n_1, n_2$ and $n_3$ can be calculated. And by introducing the hyperbolic sine function $sinh$, the relations in (30) can be put in the following form :

$$\begin{cases} \Phi_G = -\frac{g_s}{8\beta}\sum_{j=1}^{+\infty}\frac{(\xi)^j}{j[sinh(j\beta\frac{\mathcal{B}_{11}}{m})][sinh(j\beta\frac{\mathcal{B}_{22}}{m})][sinh(j\beta\frac{\mathcal{B}_{33}}{m})]} & \text{for bosons} \\ \Phi_G = \frac{g_s}{8\beta}\sum_{j=1}^{+\infty}\frac{(-\xi)^j}{j[sinh(j\beta\frac{\mathcal{B}_{11}}{m})][sinh(j\beta\frac{\mathcal{B}_{22}}{m})][sinh(j\beta\frac{\mathcal{B}_{33}}{m})]} & \text{for fermions} \end{cases} \quad (31)$$

If we introduce the $3 \times 3$ matrix $[\mathcal{B}]$ corresponding to the momenta variances, the relation (31) can be written in the following form:

$$\begin{cases} \Phi_G = -\frac{g_s}{\beta}\sum_{j=1}^{+\infty}\frac{(\xi)^j}{8j\left\{\det\left[\sinh\left(j\frac{\beta}{m}[\mathcal{B}]\right)\right]\right\}} & \text{for bosons} \\ \Phi_G = \frac{g_s}{\beta}\sum_{j=1}^{+\infty}\frac{(-\xi)^j}{8j\left\{\det\left[\sinh\left(j\frac{\beta}{m}[\mathcal{B}]\right)\right]\right\}} & \text{for fermions} \end{cases} \quad (32)$$

### 3.2 Relation between momenta variances and thermodynamic variables

In the semi-classical limit defined by the following relations (high temperature and large volume)

$$\begin{cases} \xi = e^{\beta\mu} \ll 1 \\ \beta\frac{\mathcal{B}_{ll}}{m} \ll 1 \end{cases} \quad (33)$$

the expressions of $\Phi_G$ in (31) become (at first order approximation)

$$\begin{cases} \Phi_G \cong -\frac{g_s(m)^3\xi}{8(\beta)^4\mathcal{B}_{11}\mathcal{B}_{22}\mathcal{B}_{33}} & \text{for bosons} \\ \Phi_G \cong -\frac{g_s(m)^3\xi}{8(\beta)^4\mathcal{B}_{11}\mathcal{B}_{22}\mathcal{B}_{33}} & \text{for fermions} \end{cases} \quad (34)$$

As in the reference [18, 28], we may use the QPS $(\langle\vec{p}\rangle, \langle\vec{r}\rangle)$ of a particle to deduce another approximate expression of $\Phi_G$ corresponding to the semi-classical limit (33). From the relation (28), we obtain the following approximate expression

$$\begin{cases} \Phi_G = -\frac{1}{\beta}\sum_{\{|\alpha\rangle\}}\sum_{j=1}^{+\infty}\frac{(\xi)^j e^{-j\beta\varepsilon_\alpha}}{j} \cong -\frac{g_s}{\beta}\int \xi e^{-\beta\frac{(\langle\vec{p}\rangle)^2}{2m}}\frac{d^3\langle\vec{r}\rangle d^3\langle\vec{p}\rangle}{h^3} & \text{for bosons} \\ \Phi_G = \frac{1}{\beta}\sum_{\{|\alpha\rangle\}}\sum_{j=1}^{+\infty}\frac{(-\xi)^j e^{-j\beta\varepsilon_\alpha}}{j} \cong -\frac{g_s}{\beta}\int \xi e^{-\beta\frac{(\langle\vec{p}\rangle)^2}{2m}}\frac{d^3\langle\vec{r}\rangle d^3\langle\vec{p}\rangle}{h^3} & \text{for fermions} \end{cases} \quad (35)$$

The calculation of the integral in (35) gives

$$\Phi_G \cong -\frac{g_s}{\beta}\int \xi e^{-\beta\frac{(\langle\vec{p}\rangle)^2}{2m}} \frac{d^3\langle\vec{r}\rangle d^3\langle\vec{p}\rangle}{h^3} = -\frac{g_s V}{\hbar^3}\left(\frac{m}{2\pi\beta}\right)^{\frac{3}{2}}\xi \tag{36}$$

In which $V$ is the volume containing the gas. For parallelepiped volume $= L_1L_2L_3$ , the identification between (34) and (36) gives:

$$\begin{cases} \mathcal{B}_{11} = \frac{\hbar}{2L_1}\left(\frac{2\pi m}{\beta}\right)^{1/2} \\ \mathcal{B}_{22} = \frac{\hbar}{2L_2}\left(\frac{2\pi m}{\beta}\right)^{1/2} \\ \mathcal{B}_{33} = \frac{\hbar}{2L_3}\left(\frac{2\pi m}{\beta}\right)^{1/2} \end{cases} \tag{37}$$

## 4. Thermodynamic properties

### 4.1 Thermodynamic particle number *N*

In quantum statistical mechanics, the thermodynamic particle number is not merely a counting of particles, but a key quantity that encapsulates the statistical properties of the gas under quantum constraints. Its determination involves the grand canonical potential, which reflects the interplay between particle number and the chemical potential in the quantum regime. The particle number is inherently linked to the quantum statistical properties of the system, providing a direct connection to thermodynamic quantities such as the energy and entropy. This can be formally expressed as follows:

$$N = Tr(\boldsymbol{\rho N_G}) = \sum_{\{|\mathrm{G}\rangle\}} P_G N_G = \sum_{\{|\alpha\rangle\}} \langle N_\alpha\rangle = -\frac{\partial\Phi_G}{\partial\mu} = -\beta\xi\frac{\partial\Phi_G}{\partial\xi} \tag{38}$$

From the relation (31) or (32) and (38), the calculation of the particle number gives:

$$\begin{cases} N = -\beta\xi\frac{\partial\Phi_G}{\partial\xi} = \frac{g_s}{8}\sum_{j=1}^{+\infty}\frac{(\xi)^j}{\left\{det\left[sinh\left(j\frac{\beta}{m}[\mathcal{B}]\right)\right]\right\}} & \text{for bosons} \\ N = -\beta\xi\frac{\partial\Phi_G}{\partial\xi} = -\frac{g_s}{8}\sum_{j=1}^{+\infty}\frac{(-\xi)^j}{\left\{det\left[sinh\left(j\frac{\beta}{m}[\mathcal{B}]\right)\right]\right\}} & \text{for fermions} \end{cases} \tag{39}$$

### 4.2 Internal energy

To determine the thermodynamic properties of a quantum gas, we now focus on calculating its internal energy, a fundamental quantity that encompasses the total energy of the system. Using the grand canonical ensemble, we compute this energy by evaluating the trace of the product of the density matrix and the Hamiltonian operator, which is expressed by the following relation.

$$U = Tr(\boldsymbol{\rho H_G}) = \sum_{\{|G\rangle\}} P_G E_G = \sum_{\{|\alpha\rangle\}} \varepsilon_\alpha \langle N_\alpha \rangle = -\frac{\partial(\beta\Phi_G)}{\partial\beta} + \mu N \qquad (40)$$

From the relation (31) or (32), (39) and (40), the calculation of the internal energy gives:

$$\begin{cases} U = \dfrac{g_S k_B T}{16} \displaystyle\sum_{j=1}^{+\infty} \dfrac{(\xi)^j}{det\{\left[sinh\left(j\frac{\beta}{m}[\mathcal{B}]\right)\right]\}} Tr\{\frac{\beta}{m}[\mathcal{B}]coth\left(j\frac{\beta}{m}[\mathcal{B}]\right)\} & \text{for bosons} \\ U = \dfrac{g_S k_B T}{16} \displaystyle\sum_{j=1}^{+\infty} \dfrac{-(-\xi)^j}{det\left\{\left[sinh\left(j\frac{\beta}{m}[\mathcal{B}]\right)\right]\right\}} Tr\{\frac{\beta}{m}[\mathcal{B}]coth\left(j\frac{\beta}{m}[\mathcal{B}]\right)\} & \text{for fermions} \end{cases} \qquad (41)$$

There are similarities between the expressions in (41) and the analogous results obtained for the case of the Maxwell-Boltzmann ideal gas in [28] (at high temperatures and large volumes).

$$\begin{cases} \xi = e^{\beta\mu} \ll 1 \\ \beta\dfrac{\mathcal{B}_{11}}{m} \ll 1 \end{cases} \Rightarrow \begin{cases} sinh\left(j\beta\dfrac{\mathcal{B}_{ll}}{m}\right) \cong j\beta\dfrac{\mathcal{B}_{ll}}{m} \\ coth\left(j\beta\dfrac{\mathcal{B}_{ll}}{m}\right) \cong \dfrac{m}{j\beta\mathcal{B}_{ll}} \end{cases} \qquad (42)$$

We then obtain from (41)

$$U \cong \frac{3g_s}{16}\frac{m^3\xi}{(\beta)^4\mathcal{B}_{11}\mathcal{B}_{22}\mathcal{B}_{33}} \qquad (43)$$

For both bosons and fermions, using the relations in (42), we also obtain the approximate semi-classical expression for the thermodynamic particle number $N$ as given in (39).

$$N = \frac{g_s}{8}\frac{m^3\xi}{(\beta)^3\mathcal{B}_{11}\mathcal{B}_{22}\mathcal{B}_{33}} \qquad (44)$$

It can be deduced from relations (43) and (44) that the following relation holds:

$$U = \frac{3}{2\beta}N = \frac{3}{2}Nk_BT \qquad (45)$$

The relation (45) is well known as the equation of state for a classical ideal gas.

### 4.3 Entropy

We will now compute the entropy of the system, a key quantity that characterizes the degree of disorder or randomness. For quantum systems, the entropy is defined by the Von Neumann entropy, which incorporates the quantum state distribution and the density matrix of the system. The entropy $S$ of the system is thus given by the Von Neumann entropy.

$$S = -k_B Tr[\boldsymbol{\rho}\ ln(\boldsymbol{\rho})] \qquad (46)$$

in which $\boldsymbol{\rho}$ is the density operator given in the relation (15). Using the relation (16), (38) and (40), the relation (46) can be put in the following form

$$S = -k_B \sum_{|G\rangle} P_G\, ln(P_G) = k_B ln(\Xi) + \beta k_B (U - \mu N) = -k_B \beta \Phi - k_B \beta \frac{\partial(\beta\Phi)}{\partial\beta} \quad (47)$$

By utilizing the relations (32), (39), and (41), which are derived from the quantum statistical framework, we can compute the entropy of the system. These relations provide a means to account for both the QPS effects and the statistical distribution of particles. Specifically, the entropy is derived as follows for bosons and fermions, where each term reflects the contributions from the quantum states and the specific thermodynamic parameters of the system:

$$\begin{cases} S = \frac{k_B g_s}{8} \sum_{j=1}^{+\infty} \frac{(\xi)^j}{\left\{det\left[sinh\left(j\frac{\beta}{m}[\mathcal{B}]\right)\right]\right\}} \left[\frac{1}{j} + \frac{1}{2} Tr\left\{\frac{\beta}{m}[\mathcal{B}] coth\left(j\frac{\beta}{m}[\mathcal{B}]\right)\right\} - \mu\beta\right] & \text{for bosons} \\ S = -\frac{k_B g_s}{8} \sum_{j=1}^{+\infty} \frac{(-\xi)^j}{j\left\{det\left[sinh\left(j\frac{\beta}{m}[\mathcal{B}]\right)\right]\right\}} \left[\frac{1}{j} + \frac{1}{2} Tr\left\{\frac{\beta}{m}[\mathcal{B}] coth\left(j\frac{\beta}{m}[\mathcal{B}]\right)\right\} - \mu\beta\right] & \text{for fermions} \end{cases} \quad (48)$$

### 4.4 Pressure and state equation

As in the case of the Maxwell-Boltzmann ideal gas in reference [28], the pressure becomes a matrix $[P]$ similar to the momenta variance $[B]$, and its elements are given by the following relations:

$$\begin{cases} P_{11} = -\frac{1}{L_2 L_3} \left(\frac{\partial \Phi_G}{\partial L_1}\right)_{T,\mu,L_2,L_3} \\ P_{22} = -\frac{1}{L_1 L_3} \left(\frac{\partial \Phi_G}{\partial L_2}\right)_{T,\mu,L_1,L_3} \\ P_{33} = -\frac{1}{L_1 L_2} \left(\frac{\partial \Phi_G}{\partial L_3}\right)_{T,\mu,L_1,L_2} \end{cases} \quad (49)$$

By applying the relations in (31), (32), which are derived from the quantum statistical framework and (49), we can derive the explicit expressions for the pressure matrices as follows.

$$\begin{cases} [P] = \frac{g_S k_B T}{8V} \sum_{j=1}^{+\infty} \frac{(\xi)^j}{det\{[sinh\left(j\frac{\beta}{m}[\mathcal{B}]\right)]\}} \left[\frac{\beta}{m}[\mathcal{B}] coth\left(j\frac{\beta}{m}[\mathcal{B}]\right)\right] & \text{for bosons} \\ [P] = \frac{g_S k_B T}{8V} \sum_{j=1}^{+\infty} \frac{-(-\xi)^j}{det\{[sinh\left(j\frac{\beta}{m}[\mathcal{B}]\right)]\}} \left[\frac{\beta}{m}[\mathcal{B}] coth\left(j\frac{\beta}{m}[\mathcal{B}]\right)\right] & \text{for fermions} \end{cases} \quad (50)$$

It follows that we have the following equations of state (which are matrix equations like in [28])

$$\begin{cases} [P]V = \dfrac{g_S k_B T}{8} \displaystyle\sum_{j=1}^{+\infty} \dfrac{(\xi)^j}{det\{\ [sinh\left(j\dfrac{\beta}{m}[\mathcal{B}]\right)]\}} \left[\dfrac{\beta}{m}[\mathcal{B}]\ coth\left(j\dfrac{\beta}{m}[\mathcal{B}]\right)\right] & \text{for bosons} \\ [P]V = \dfrac{g_S k_B T}{8} \displaystyle\sum_{j=1}^{+\infty} \dfrac{-(-\xi)^j}{det\{\ [sinh\left(j\dfrac{\beta}{m}[\mathcal{B}]\right)]\}} \left[\dfrac{\beta}{m}[\mathcal{B}]coth\left(j\dfrac{\beta}{m}[\mathcal{B}]\right)\right] & \text{for fermions} \end{cases} \quad (51)$$

In the semi-classical limit, as defined by the relations in (42), it can be deduced from equations (43), (44), and (51) that this difference between bosons and fermions vanishes, the pressure becomes a scalar quantity, and we recover the well-known classical equation of state for an ideal gas.

$$PV = Nk_B T \quad (52)$$

## 5. Discussions

The improved models of ideal Fermi and Bose gases developed in this work provide significant advancements in quantum statistical mechanics by incorporating QPS effects, including geometric constraints such as shape and size. These effects are crucial for accurately describing quantum gases at low temperatures and small volumes, where quantum mechanical constraints become more pronounced.

Thermodynamic Particle Number (*N*): The thermodynamic particle number *N*, given in equation (39), is calculated using discrete summations rather than integrals, reflecting the quantum statistical properties of the system. These discrete sums are influenced by the quantum variances of the momenta, and the quantum shape and size effects are embedded in the parameters that describe the geometry of the system, such as the volume and shape of the container. The determinant of a matrix, as it appears in the expression of the grand canonical potential (linked to equation (37)), plays a key role in demonstrating these effects. Momentum variance matrices provide a detailed description of how quantum fluctuations influence particle distribution. The quantum shape and size of the system are encoded in the determinant of matrices, such as those in equations (49) and (50), which link the pressure and thermodynamic particle number to the spatial geometry of the system. The presence of matrices in these expressions explicitly shows how the geometry and size of the system influence the thermodynamic quantities. This reflects the underlying quantum statistical nature of the gas.

Internal Energy (*U*): The internal energy *U*, expressed in equation (41), is also determined through discrete summations. The latter involves calculating the determinant and trace of a matrix, so the internal energy is affected by the quantum statistical properties of the system, including the quantum shape and size effects. These quantum effects are encoded in the variances of the momenta, which in turn are influenced by the geometry of the system. The influence of shape and size is evident through the use of matrices that involve determinants, such as in the grand canonical potential (equation (37)), and their contribution to the calculation of thermodynamic quantities like the internal energy. The equation of internal energy (45) for a classical ideal gas is recovered at high temperatures and large volumes according to the relation (42).

Entropy (*S*): The entropy calculation, shown in equation (48), uses QPS effects, which are influenced by the quantum shape and size of the system. The entropy expression also requires the calculation of a determinant and a trace. The matrix expressions inside the equation allows for a more detailed understanding of how entropy is impacted by the geometry of the system.

The shape and size effects are embedded in the momentum variance matrices, which modify the entropy formula.

Pressure and State Equation: The pressure, calculated in equation (50), is expressed in terms of momentum variance matrices, which capture the quantum shape and size effects. These matrices, derived from the grand canonical potential and the density of states, show how the geometry of the system influences the pressure. The quantum shape and size effects are reflected in the determinant of the matrices and the momentum distribution, as seen in the matrix expressions for pressure in equations (49) and (50). In the semi-classical limit, as shown in equations (43) and (44), the quantum effects decreased, and the model recovers the classical equation. The same influence is observed in the pressure. Furthermore, the difference between bosons and fermions vanishes, and the pressure becomes a scalar quantity, which reduces to the classical equation of the Maxwell-Boltzmann ideal gas (52). However, at lower temperatures and smaller volumes, these quantum shape and size effects are essential and are fully captured in the matrix form of the thermodynamic quantities. The emergence of a diagonal pressure matrix [P], instead of a scalar quantity, has a clear physical interpretation. It indicates that under confinement, the momentum variances along different spatial directions may not be identical, leading to possible anisotropic pressure components. In nanoscale or subatomic systems, such anisotropy reflects directional dependence of quantum stress due to boundary constraints and quantization of motion. Each diagonal component of [P] can thus be interpreted as a partial pressure along a specific axis, corresponding to distinct confinement lengths or quantization scales. This formulation reveals that confinement does not violate the fundamental isotropy of physical laws but rather modifies their manifestation through boundary-induced anisotropy. The isotropic classical pressure $P = Nk_BT/V$ is recovered in the semiclassical limit, where the confinement effects vanish and all diagonal elements of $[P]$ become equal. This highlights that the diagonal pressure matrix $[P]$ provides a generalized tensorial description of quantum pressure, bridging microscopic directional stresses with macroscopic thermodynamic pressure.

In summary, the matrix formulations in equations (37), (49), and (50) directly represent how quantum shape and size effects influence the thermodynamic properties of ideal gases. These expressions encode the geometry of the system and momentum variances, providing a more detailed description of quantum statistical behaviors not apprehended by standard models. The use of matrices in the pressure and equations of state further highlights how the quantum mechanical description, including the spatial geometry, impacts thermodynamic properties like entropy, reflecting quantum fluctuations and shape-dependent behavior.

## 6. Conclusions

This work contributes to improvements in the Fermi-Dirac and Bose-Einstein ideal gas models by incorporating the effects of the quantum nature of phase space, which are inherently related to the uncertainty principle. These corrections, arising from quantum phase-space effects, are expected to be significant at low temperatures and in confined volumes. They allow for revealing the existence of quantum shape and size effects, as shown, for instance, by the equations of state (51).

The improved models of ideal Fermi and Bose gases, established in this work, are naturally compatible with the uncertainty principle. These models allow for new explicit expressions for the thermodynamic properties of the ideal gases. For instance, we have the relations in (32) for the grand canonical potential, (39) for the thermodynamic particle number, (41) for the internal energy, (48) for the entropy, and (50) for the pressure. In the standard models of the ideal Fermi gas and ideal Bose gas, these thermodynamic quantities are expressed as integrals, while in the present improved models, they are expressed as discrete summations. These expressions from the improved models explicitly show that the thermodynamic quantities are influenced by the quantum statistical variance of momentum, highlighting the role of quantum fluctuations in the system. These variances are also related to the thermodynamic variables, such as temperature, shape and size of the volume, as shown in equation (37). It was also shown that the well-known relations corresponding to the classical ideal gas model can be recovered as asymptotic limits at high temperatures and large volumes.

The results obtained in this work are likely to find useful applications in areas such as condensed matter physics, astrophysics, and the study of quantum fluids, where the ideal Fermi gas and ideal Bose gas models are commonly used. Some quantum corrections typically arise from the bosonic or fermionic nature of the particles involved [6, 30-33]. However, additional quantum corrections due to confinement and geometric effects can become significant in certain cases, such as in nanoscience and nanotechnology [11-12, 34-36]. According to the present study, these additional corrections can be described as arising from the quantum nature of phase space. Thus, this work is expected to have important implications in the fields of nanotechnology and quantum computing. A potential connection between this work and recent research lies in the improvements made to quantum gas models, incorporating low-temperature quantum effects and confined volumes, could provide a theoretical foundation for exploring shape-induced Bose-Einstein [37]. The extension of these results to the modeling of real gases could lead to significant advancements in our understanding and applications of such systems. Finally, the potential introduction of relativistic corrections, as discussed in [29], may also require further consideration.